\documentclass[journal,twoside,web]{ieeecolor}
\providecommand{\refname}{References}
\usepackage{generic}
\usepackage{cite}
\usepackage{amsmath,amssymb,amsfonts}
\usepackage{algorithmic}
\usepackage{graphicx}
\usepackage{algorithm,algorithmic}
\usepackage{hyperref}
\hypersetup{hidelinks}
\usepackage[normalem]{ulem}
\usepackage{multirow}
\usepackage{xcolor}
\usepackage{textcomp}
\usepackage{algorithm}
\usepackage{algorithmic}

\definecolor{JamesBlue}{rgb}{0.0, 0.0, 0.0}

\def\BibTeX{{\rm B\kern-.05em{\sc i\kern-.025em b}\kern-.08em
    T\kern-.1667em\lower.7ex\hbox{E}\kern-.125emX}}
\begin{document}

\title{Decoding Covert Speech from EEG Using a Functional Areas Spatio-Temporal Transformer}
\author{Muyun Jiang, Yi Ding, Wei Zhang, Kok Ann Colin Teo, LaiGuan Fong, Shuailei Zhang, Zhiwei Guo, Chenyu Liu, Raghavan Bhuvanakantham, Wei Khang Jeremy Sim, Chuan Huat Vince Foo, Rong Hui Jonathan Chua, Parasuraman Padmanabhan, Victoria Leong, Jia Lu, Balázs Gulyás, Cuntai Guan \IEEEmembership{Fellow, IEEE}
\thanks{Muyun Jiang, Yi Ding, Shuailei Zhang, Zhiwei Guo, Chenyu Liu and Cuntai Guan are with the College of Computing and Data Science, Nanyang Technological University, Singapore.}
\thanks{Chuan Huat Vince Foo and Rong Hui Jonathan Chua are with DSO National Laboratories, Singapore.}
\thanks{Wei Zhang, Kok Ann Colin Teo, LaiGuan Fong, Raghavan Bhuvanakantham, Wei Khang Jeremy Sim, Parasuraman Padmanabhan, and Balázs Gulyás are with the Cognitive Neuroimaging Centre and Lee Kong Chian School of Medicine, Nanyang Technological University, Singapore.}
\thanks{Kok Ann Colin Teo is also with the Division of Neurosurgery, National University Health System, Singapore, and the IGP-Neuroscience, Interdisciplinary Graduate Programme, Nanyang Technological University, Singapore.}
\thanks{Victoria Leong is with the Division of Psychology, Nanyang Technological University, Singapore, and the Department of Pediatrics, University of Cambridge, United Kingdom.}
\thanks{Jia Lu is with the Yong Loo Lin School of Medicine, National University of Singapore, Singapore.}
\thanks{Balázs Gulyás is also with the Department of Clinical Neuroscience, Karolinska Institutet, Stockholm, Sweden.}
\thanks{Cuntai Guan is the Corresponding Author. E-mail: ctguan@ntu.edu.sg}
}
\maketitle

\begin{abstract}

Covert speech involves imagining speaking without audible sound or any movements. Decoding covert speech from electroencephalogram (EEG) is challenging due to a limited understanding of neural pronunciation mapping and the low signal-to-noise ratio of the signal. In this study, we developed a large-scale multi-utterance speech EEG dataset from 57 right-handed native English-speaking subjects, each performing covert and overt speech tasks by repeating the same word in five utterances within a ten-second duration. Given the spatio-temporal nature of the neural activation process during speech pronunciation, we developed a Functional Areas Spatio-temporal Transformer (FAST), an effective framework for converting EEG signals into tokens and utilizing transformer architecture for sequence encoding. Our results reveal distinct and interpretable speech neural features by the visualization of FAST-generated activation maps across frontal and temporal brain regions with each word being covertly spoken, providing new insights into the discriminative features of the neural representation of covert speech. This is the first report of such a study, which provides interpretable evidence for speech decoding from EEG. The code for this work has been made public at \url{https://github.com/Jiang-Muyun/FAST}

\end{abstract}


\section{Introduction}

Covert speech is the process of internally articulating speech units, such as words or sentences, without producing audible sounds or movements \cite{metzger2024highly}. This method is particularly beneficial for individuals with verbal communication impairments due to conditions like stroke, trauma, and amyotrophic lateral sclerosis (ALS), which affect speech production and comprehension \cite{metzger2023high, silva2024speech}. This task also helps provide insights into how the brain forms speech patterns and prepares them for verbal articulation, even when no physical speech occurs. Decoding covert speech has made tremendous progress in using invasive recordings \cite{angrick2019speech, lu2023neural}. Decoding from EEG has gained recognition as a method for assisting individuals with severe communication barriers to establish a channel for thought-based interaction \cite{gu2021eeg}. However, the task of decoding covert speech from non-invasive neural recordings such as EEG remains challenging.

Currently, most covert speech BCIs are based on invasive methods, their feasibility has been demonstrated through various studies. This includes utilizing electrocorticography (ECoG) data to synthesize speech using densely connected 3D CNN as shown in \cite{angrick2019speech}. Moreover, \cite{duraivel2023high} demonstrated that the finer details captured by these recordings play a crucial role in understanding and interpreting speech-related neural activity. Additionally, \cite{proix2022imagined} has shown that imagined speech could be decoded from low and cross-frequency features in intracranial electroencephalography (iEEG) signals. Non-invasive technologies have gradually garnered attention for their potential in BCIs. Notably, the use of Magnetoencephalography, as explored in \cite{dash2020decoding}, has successfully decoded imagined and spoken phrases. Despite these advancements, studies based on EEG remain scarce due to the challenge of small signal magnitudes, which hinder the ability of decoding methods to learn effective features.

In recent years, with the incorporation of deep learning technologies, particularly CNNs like DeepConvNet \cite{schirrmeister2017deep} and EEGNet \cite{lawhern2018eegnet}, the capability to accurately interpret EEG data has markedly improved in the tasks of motor imagery, emotional cognition, and more \cite{thomas2008adaptive, mahamune2021classification, rahman2021emotion, ding2023lggnet, ieracitano2020novel, power2011functional, song2018eeg, zhong2020eeg, klepl2022eeg}. Building on the success of the transformer model in the fields of computer vision and natural language processing, researchers have also attempted to adopt the transformer structure for the identification and classification of brain activities, leveraging their ability to effectively model long sequences \cite{parmar2018image, xie2022transformer, song2022eeg, bagchi2022eeg}. However, challenges persist in effectively integrating transformer-based architectures with EEG data. Issues such as high computational demand, the need for large datasets to train effectively, and the sensitivity to the highly variable nature of EEG signals complicate the application. 

From the existing literature, numerous studies have focused on EEG classification tasks using pure convolutions or convolutional layers combined with transformer architecture approaches. However, a comprehensive insight into the limiting factors of these approaches reveals a common set of challenges:

\begin{enumerate}
    \item Lack of a high-quality large-scale, multi-utterance, EEG-based covert speech dataset.
    \item  Lack of effective decoding algorithm to tackle challenging covert speech EEG signals.
    \item Lack of understanding of the covert speech intrinsic information in EEG to enable speech decoding.
\end{enumerate}

Addressing the challenges outlined, we have collected a multi-utterance dataset aimed at exploring the brain's mechanisms during covert speech activities. This dataset is collected from 57 right-handed adult males who engaged in both covert and overt speech tasks. Each participant contributed 1000 utterances, allowing for a detailed analysis of speech processing and brain function. 

\textcolor{JamesBlue}{
% We have designed FAST, a Functional Areas Spatio-temporal Transformer model that combines a convolutional neural network with a transformer architecture. This model initiates with a serial of brain functional area convolutional tokenizers, specifically designed to capture spatially sensitive features from each data. Tokens from each brain region first pass through a spatial projection layer to integrate information from various brain regions. Subsequently, these aggregated features undergo deeper analysis through multiple transformer layers, enhancing our ability to decode and interpret complex brain signals. Detailed feature visualization uncovers distinct and interpretable speech-neural patterns across frontal and temporal brain regions for covertly spoken words, offering new insights into the discriminative features underlying the neural representation of covert speech.
We developed FAST, a Functional Areas Spatio-temporal Transformer model that integrates a convolutional neural network with a transformer architecture. The model begins with a series of brain functional area convolutional tokenizers designed to extract spatially sensitive features from the data. Tokens from each brain region are first processed through a spatial projection layer to integrate information across different brain regions. These aggregated features then undergo deeper analysis through multiple transformer layers, enhancing the model’s ability to decode and interpret complex brain signals. Detailed feature visualization reveals distinct and interpretable speech-neural patterns in the frontal and temporal brain regions for covertly spoken words, providing new insights into the discriminative features underlying the neural representation of covert speech.
}

Our main contributions can be summarized as follows:

\begin{enumerate}
    \item Proposed FAST, a novel Spatio-temporal Transformer Network inspired by cerebral functional systems for covert speech EEG modeling.
    \item Provided in-depth feature analysis, revealed discriminative features during covert speech.
    \item \textcolor{JamesBlue}{We evaluate the model on a large-scale multi-utterance covert and overt speech dataset from 57 subjects totaling 1000 utterances per participant.}
\end{enumerate}

To the best of the authors' knowledge, this is the first study in the literature showing features learned from EEG-based covert speech BCI approaches. This model merges neuroscience-driven feature extraction with a multimodal dataset and a flexible learning strategy. 

The structure of this work is outlined as follows: Section \ref{section:Related_Work} reviews prior research closely related to our study. Section \ref{section:Methodology} elaborates on the detailed methodology of our model. Section \ref{section:Dataset} focuses on the procedures for data collection. Section \ref{section:Experiment_Settings} expressed the training scheme of FAST. Section \ref{section:Results_Findings} outlines the computational experimental setup and the evaluation of our novel method as well as our insights derived from in-depth feature analysis. Finally, Section \ref{section:Conclusion} concludes the study and proposes directions for future research.

\section{Related Work}
\label{section:Related_Work}

\subsection{Neural Foundations and Decoding Advancements of Covert Speech}

Recent studies provide neural evidence that covert speech involves an articulatory component \cite{lu2023common}.  Activation in Broca's area, associated with speech production, and supplementary motor areas underscores its link to articulatory processes, while engagement of auditory imagery networks highlights its reliance on both auditory and motor functions \cite{fernyhough2023inner, tian2016mental}. Research has identified distinct neural signatures for covert speech compared to listening to speech, with overlapping yet specialized activation in key brain regions \cite{nalborczyk2023distinct}.

\textcolor{JamesBlue}{
Recent advancements in neural decoding of imagined speech have significantly contributed to our understanding of brain processes. For instance, one study \cite{kim2023diff} employs a diffusion-based model to decode imagined speech from EEG signals, demonstrating the potential of machine learning in enhancing speech recognition accuracy. Another study \cite{lee2023speech} investigates speech synthesis from brain signals using a generative model, paving the way for neural activity-based communication devices. Additionally, \cite{guo2025enhancing} explores transfer learning from overt to imagined speech through a convolutional autoencoder, improving EEG-based imagined speech decoding. A separate framework \cite{lee2023towards} integrates spoken speech data to refine neural decoding models, providing a more direct interpretation of internal speech representations. Moreover, \cite{ramirez2023novel} introduces a deep capsule neural network to decode vowel imagery from EEG, improving the efficiency of imagined speech processing.}

\textcolor{JamesBlue}{Despite these advancements, existing studies often fall short in providing detailed insights into the neurological mechanisms underlying covert speech and lack fine-grained temporal visualizations of neural features. A deeper exploration of neural substrates could bridge these gaps, offering a clearer understanding of covert speech processes and improving BCI design for more effective and intuitive covert speech recognition.
}

\subsection{Transformer models for EEG}

Building upon the success of transformer models in computer vision and natural language processing, researchers have explored their application in EEG analysis to capture complex spatio-temporal patterns inherent in brain activity. For instance, \cite{song2022eeg} proposed a transformer-based EEG Conformer that effectively learns spatial and temporal features from EEG data, achieving state-of-the-art performance in emotion recognition tasks. Recently, \cite{ding2024masa} introduced a multi-anchor space-aware temporal transformer model designed for EEG decoding, demonstrating its efficacy across various datasets. Additionally, \cite{ding2024emt} developed a transformer architecture that analyzes spatio-temporal dynamics of EEG signals to estimate emotion states.

\textcolor{JamesBlue}{However, these studies often lack clear visualizations of the tokenization process, making it challenging to interpret and transfer the learned temporal tokens effectively. Furthermore, there is limited investigation into how these temporal tokens interact with spatial features to capture the intricate dynamics of EEG signals. A more comprehensive exploration of this interplay could provide deeper insights into the spatiotemporal characteristics of brain activity to enhance the interpretability and robustness of neural decoding models.}

\section{Methodology}
\label{section:Methodology}

\begin{figure*}[!ht]
\includegraphics[width=\textwidth]{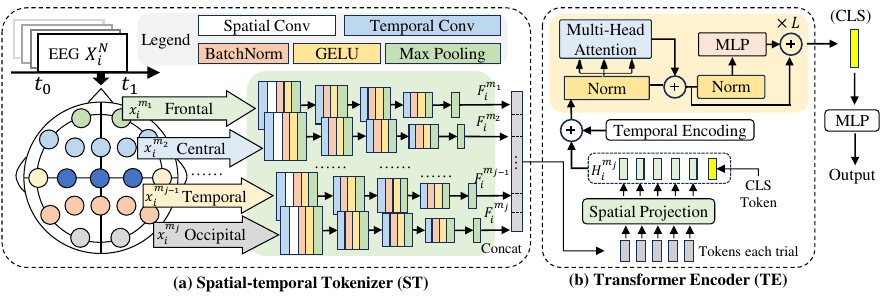}
\caption{Overview of proposed FAST. (a) The Spatial-temporal Tokenizer (ST) block illustrates the initial processing of EEG data through spatial and temporal convolutional layers. (b) The Transformer Encoder (TE) block shows the transformer architecture used for tokens generated by ST, which outputs a learned CLS token for classification results.}
\label{fig:Structure}
\end{figure*}

In this section, we introduce the model structure of FAST. As depicted in Figure \ref{fig:Structure}, FAST comprises two primary components: the Spatial-temporal Tokenizer (ST) and the Transformer Encoder (TE) Block. ST is responsible for processing information from each brain function area, while the TE blocks utilize a transformer architecture to operate on the tokens generated by ST. The network undergoes a two-step pre-training process followed by a fine-tuning process. The source code of this work has been opened at \url{https://github.com/Jiang-Muyun/FAST}

\subsection{Spatial-temporal Tokenizer}

\begin{figure}[!ht]
\centering
\includegraphics[width=0.85\linewidth]{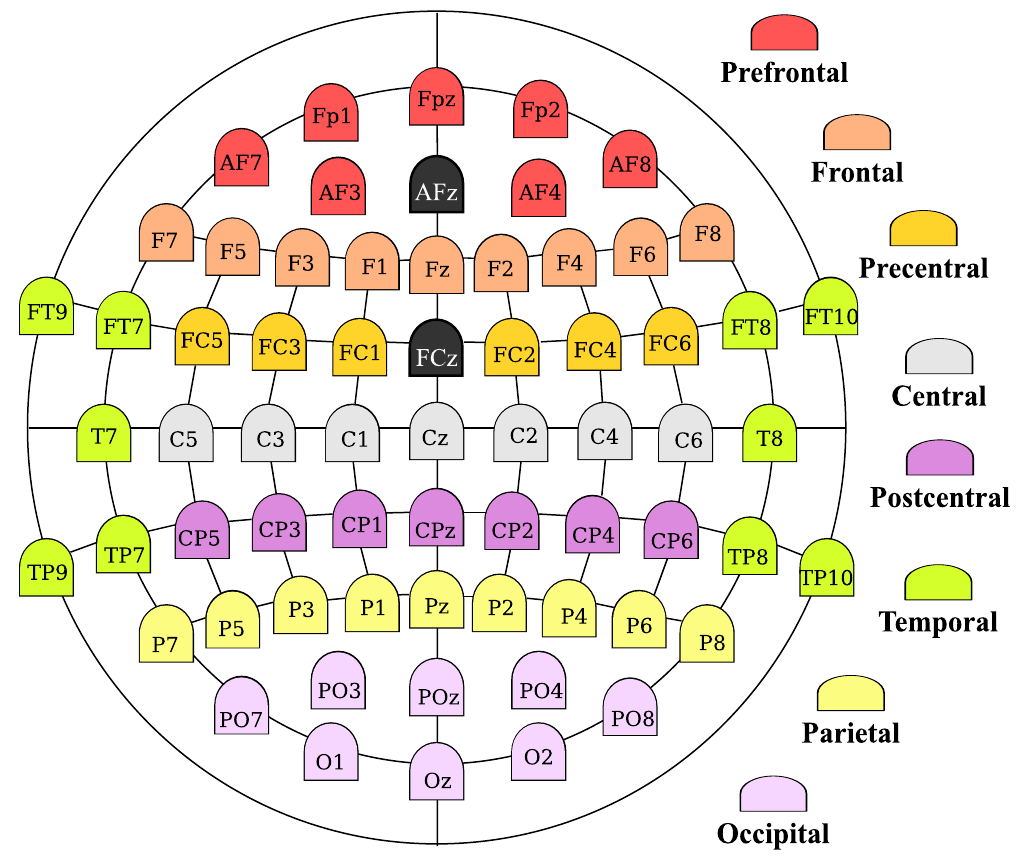}
\caption{The divided brain functional regions are based on the spatial locations of EEG channels, where FCz serves as the reference electrode, and AFz serves as the ground electrode.}
\label{fig:EEG_Zoned}
\end{figure}

The ST block is composed of multiple independent brain functional area encoders, each responsible for processing information only corresponding to specific brain functions. Given an EEG trial input with $N$ channels, denoted as $X^{N}$, our preliminary step involves segmenting the EEG trial into shorter segments. This segmentation is achieved by applying a fixed-size window and slide across the trial, ultimately resulting in $S$ segments. Each segment, particularly the $i^{th}$ segment, is denoted as $X_{i}^{N}$, where $i$ ranges from $1$ to $S$ indicating the segment index. Following this segmentation, the next step is to partition the EEG signals into $M$ distinct brain areas. 

The brain's functions are complex networks with hierarchical organization across neurons, local circuits, and functional areas \cite{power2011functional, kober2008functional, allen2018frontal}. EEG electrodes are positioned on the scalp following the 10-10 \cite{koessler2009automated} system, which organizes channels based on their placement over distinct regions of the brain. In this study, we segment the brain into different functional regions according to the spatial locations of EEG channels, similar to LGGNet \cite{ding2023lggnet}, as illustrated in Figure \ref{fig:EEG_Zoned}.

This division allows us to examine the specific roles of various brain regions, particularly focusing on the functional responsibilities of the frontal \cite{allen2018frontal} and temporal regions \cite{oliveri2009spatial}. Specifically, the frontal region is associated with executive functions, decision-making, and attentional control, while the temporal region plays a critical role in auditory processing, memory, and language comprehension. By isolating these areas through the 10-10 electrode system, the network will be able to learn and capture region-specific spatial features more effectively \cite{ding2023lggnet}.

Mathematically, the division can be represented as a partitioning process
\begin{equation}
P: X_{i}^{N} \rightarrow \{x_{i}^{m_1}, x_{i}^{m_2}, \ldots, x_{i}^{m_j}\}
\end{equation}

\noindent where $P$ signifies the partitioning function that takes a subset of channels from each segment $X_{i}^{N}$ to a set of $M$ brain areas for that segment, and $x_{i}^{m_j}$ denotes the subset of channels related to the $j^{th}$ brain area in the $i^{th}$ segment.

Each partitioned brain area $x_{i}^{m_j}$, undergoes processing through a specific functional area encoder $f_j(x)$, a sequence of convolutional layers, normalization layers, activation functions, and max pooling operations. A spatial-temporal module is placed at the beginning of the series of convolutions to capture the spatial-temporal information in each functional area. Denoted as $\text{Conv}_T$ and $\text{Conv}_S$. This sequential processing is represented as follows:
\begin{equation}
x_{i, \text{temporal}}^{m_j} = \text{Conv}_T\left(x_{i}^{m_j}\right)
\end{equation}
\begin{equation}
x_{i, \text{spatial}}^{m_j} = \text{Conv}_S\left(x_{i, \text{temporal}}^{m_j}\right)
\end{equation}
After extracting spatial and temporal features, the data is further processed through Batch Normalization (BN), Gaussian Error Linear Units (GELU) activation functions, and max pooling operations. This is expressed as:

\begin{equation}
x_{i(1)}^{m_j} = \text{MaxPool}\left(\text{GELU}\left(\text{BN}\left(x_{i, \text{temporal}}^{m_j}\right)\right)\right)
\end{equation}

where $\Phi(x)$ is the cumulative distribution function of the standard Gaussian distribution.

After the first spatial-temporal encoder, a series of temporal encoders are applied to $x^{m_j}_{i}$ to extract the short-time information within each functional area. The temporal encoder comprises convolutional layers, normalization layers, activation functions, and pooling operations. The process is outlined as follows: $l \in [2, L_t]$ and $L_t$ is the total layer number of the temporal convolution layers:
\begin{equation}
x_{i(l)}^{m_j} = \text{MaxPool}\left(\text{GELU}\left(\text{BN}\left(\text{ConvT}_l\left(x^{m_j}_{i(l-1)}\right)\right)\right)\right) 
\end{equation}

To achieve a representation of the spatial signals that is invariant to the length of the EEG segments, global max pooling is applied across the time dimension, transforming the signal into fixed-dimensional vectors, each representing the most significant signal peak within the duration for its respective brain area:
\begin{equation}
F_{i}^{m_j} = \text{GlobalMaxPool}_{\text{time}}\left(x^{m_j}_{i(L_t)}\right)
\end{equation}

$F_{i}^{m_j}$ represent the feature vector of the $j^{th}$ brain functional area of the $i^{th}$ segment. 

\subsection{Transformer Encoder}

The TE blocks utilize a transformer architecture that operates on the tokens generated by ST, by first going through spatial projection layers and then transformer blocks. The spatial projection layers operate on the spatial dimension, while transformer blocks operate on the temporal dimension.

The spatial projection involves processing through a series of $L_s$ layers of multi-head attention mechanisms, each focusing on refining the representation of individual brain functional areas. Formally, for the $j^{th}$ brain functional area in the $i^{th}$ segment, the process can be expressed as follows
\begin{equation}
H^{m_j}_{i,(0)} = F_{i}^{m_j},
\end{equation}

The initial feature vector \( H^{m_j}_{i,(0)} \) serves as the input for the first layer. In subsequent layers, the output from the previous layer undergoes a transformation via the multi-head attention function \( \text{Multi-Head}(\cdot) \). The transformation at each layer $l$ is defined as:
\begin{equation}
H^{m_j}_{i,(l)} = \text{Multi-Head}(H^{m_j}_{i,(l-1)}, H^{m_j}_{i,(l-1)}, H^{m_j}_{i,(l-1)})
\end{equation}
\begin{equation}
\delta = \text{Multi-Head}(Q, K, V) = \text{Concat}(head_1, \dots, head_h) W^O
\end{equation}
\begin{equation}
head_h = \text{Attention}(QW_h^Q, KW_h^K, VW_h^V)
\end{equation}

where $Q$, $K$, and $V$ represent the query, key, and value matrices, $W_h^Q$, $ W_h^K$, and $W_h^V$ are the weight matrices specific to each head for the query, key, and value. The outputs from individual heads are concatenated and then transformed by multiplying with the output weight matrix $W^O$. 

After each multi-head attention layer, a position-wise feed-forward network (FFN) is added to form a transformer layer. The FFN is formulated as:
\begin{equation}
\text{FFN}(\delta) = W_2\text{GELU}(W_1 \delta + b_1) + b_2
\end{equation}
\begin{equation}
\delta' = \text{LN}(\delta + \text{FFN}(\delta))
\end{equation}
where $W_1$, $W_2$ are the weight matrices and $b_1$, $b_2$ are the biases for the FFN. Layer normalization (LN) and residual connections are applied at the end of FFN, where $\delta$, the output of the Multi-Head Attention, serves as the input to the FFN, and $\delta'$ represents the output of the FFN.

Then, $L_s$ transformer layers in the spatial projection block are applied iteratively to process the data from each brain region, forming an output of $H^{m_j}_{i,(L_s)}$. By concatenating them for each region $j \in \{1, \dots, J\}$ into a single feature vector, we obtain $G_i$, a refined representation of all the brain functional areas within the $i^{th}$ segment. This process can be expressed as:
\begin{equation}
G_i = \bigoplus_{j=1}^J H^{m_j}_{i(L_{sp})} 
\end{equation}
where $\bigoplus $ denotes the concatenation operation. This results in a comprehensive feature vector $ G_i $ that encapsulates the functional dynamics of all the brain regions for the $ i^{th} $ segment.

Following spatial projection, a transformer block comprising \(L\) layers, which leverages self-attention and FFNs as previously described, is employed on the sequence of extracted feature vectors \(G_i\) to decode temporal information across trials. To better capture temporal dynamics within the transformer framework, a temporal encoding, denoted by \(T_i\), is introduced to each input feature vector. This encoding is initially randomized and subsequently learned during training. Additionally, a classification token represented as \(F_{\text{cls}}\) is introduced and similarly learned during training. The input tokens for this transformer are presented as follows:
\begin{equation}
G = \{G_1 + T_1, G_2 + T_2, \ldots, G_i + T_i, F_{\text{cls}} + T_{i+1}\},
\end{equation}
The transformer operates iteratively as depicted:
\begin{equation}
G^L = \text{Transformer}_{l}(G^{l-1})
\end{equation}
for \(l \in \{1, \ldots, L\}\). Each transformer block applies self-attention across the entire sequence of feature vectors, including the classification token, facilitating interactions among various segments of the input. The final classification token, \(G^L_{\text{cls}}\), is derived after multiple layers of attention and subsequently processed through a classification head for predictions:
\begin{equation}
\hat{y} = \text{MLP}(G^L_{\text{cls}})
\end{equation}

where $\hat{y}$ represents the predicted probabilities, with a shape of (B,5) corresponding to the 5 classes in our classification task. The algorithm detailed above, which we will refer to as Algorithm \ref{algo1}, can be briefly summarized as follows:

\begin{algorithm}[]
\caption{Data Processing and Feature Extraction}
\label{algo1}
\begin{algorithmic}
\REQUIRE EEG input $X^N$ with $N$ channels
\STATE Segment $X^N$ into $S$ segments $X_i^N$, $i=1$ to $S$
\STATE Partition $X_i^N$ into $M$ areas: $P: X_i^N \rightarrow \{x_i^{m_1}, \ldots, x_i^{m_M}\}$
\FOR{each area $m_j$ in segment $i$}
    \STATE Encode $x_i^{m_j}$ using $f_j(x)$
    \STATE Apply spatial-temporal processing to $x_i^{m_j}$
    \STATE Normalize, activate, and pool: $x_{i(1)}^{m_j}$
    \STATE Process for invariant representation: $x_{i(L_{t})}^{m_j}$
    \STATE Global max pooling: $F_{i}^{m_j}$
\ENDFOR
\STATE Apply spatial projection to get $H^{m_j}_{i,(L_s)}$
\STATE Concatenate to get $G_i$ for each segment
\STATE Prepare sequence $G$ with $G_i$ and $G_{\text{cls}}$
\STATE Encode sequence with transformer
\STATE Predict $\hat{y} = \text{MLP}(G^L_{cls})$
\STATE Return $\hat{y}$
\end{algorithmic}
\end{algorithm}

\section{Dataset}
\label{section:Dataset}

\begin{figure*}[]
\includegraphics[width=\textwidth]{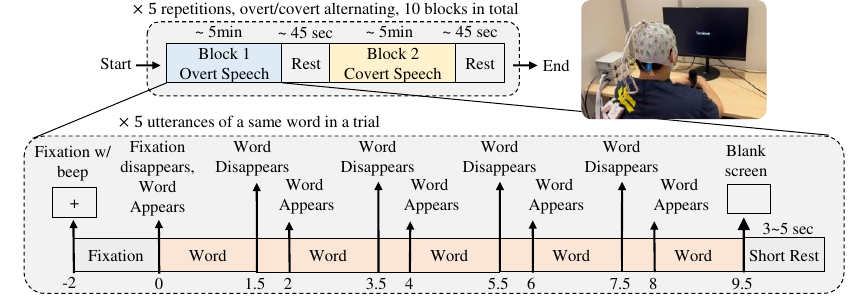}
\caption{Protocol of the experiment. (Top): The experiment is structured into blocks. Each subject will complete 10 blocks, alternating between overt and covert EEG experiments. Each block consists of 20 trials, presenting five words in a pseudo-random order. (Bottom): In each trial, the same words were displayed on the screen at predetermined intervals (T = 0, 2, 4, 6, 8 seconds) and vanished at T + 1.5 seconds. Subjects are instructed to overtly pronounce or covertly imagine pronouncing the words five times following the blink of the words. A brief resting period, with a random duration ranging from 3 to 5 seconds, is provided after each trial.}
\label{fig:Protocal}
\end{figure*}

\subsection{Protocol Settings}

We collected a multi-utterance dataset during overt and covert speech activities. The dataset comprises recordings from 57 right-handed native English-speaking adult males, each performing covert and overt speech tasks by repeating the same word in five utterances within a ten-second duration, with a total of 1000 utterances per individual. The average age of the subjects was 24.2 ± 3.6 years. The information of the collected dataset is shown in Table \ref{tab:experiment_settings}. The data collection adhered to the guidelines of the Declaration of Helsinki and took place at the Cognitive Neuroimaging Centre (CoNiC) at Nanyang Technological University (NTU) in Singapore. Written informed consent was obtained from all participants. The study received ethical approval from the Institutional Review Board (IRB) of NTU under the approval number IRB-2022-040.

\begin{table}[ht]
\centering
\caption{Summary of Data Collection Settings}
\begin{tabular}{ll}
\hline
\textbf{Feature}                     & \textbf{Details}\\
\hline
Subjects                            & 57\\
Modality                            & EEG\\
Experiment Type                     & Overt and covert and word-speaking\\
Total Blocks                        & 10 (5 blocks for each condition)\\
Block Duration                      & 5.5 minutes\\
Rest Between Blocks                 & 45 seconds\\
Trials per Block                    & 20\\
Words per Trial                     & 5 utterances\\
Selected Words                      & "Go there", "Distract Target", \\
                                    & "Follow me", "Explore Here", "Terminate" \\
\hline
\end{tabular}
\label{tab:experiment_settings}
\end{table}

As shown in Fig \ref{fig:Protocal}, the experimental protocol comprises a total of 10 blocks, including alternating overt speech blocks and covert speech blocks, totaling 5 blocks for each condition. Every block lasts about 5.5 minutes, followed by a short rest of 45 seconds. There are a total of 20 trials in each block, participants were instructed to speak overtly or covertly, 5 repetitions in each 10-second trial. Between trials, the words were presented in a pseudo-random order to reduce predictability. During the overt speech blocks, participants were directed to speak each word loudly and clearly, keeping the natural tone of their daily speech. During the covert speech blocks, participants were instructed to imagine the pronunciation of the word without any physical movement or production of any audible sound. We chose to collect a multi-utterance dataset to ensure more stable and robust neural responses. The single utterance is prone to variability due to factors like attention lapses or insufficient context, which can adversely affect the decoding and induction of cognitive signals. 

We used 5 words for speaking in this experiment based on the following criteria: 1) Adequate length featuring a polysyllabic structure.
2) Distinct articulation between terms.
3) To enhance the system's applicability for BCI, the chosen words should relate to robot control functions. The chosen words are "Go there", "Distract Target", "Follow me", "Explore Here", and "Terminate".

\subsection{Data Collection and Preprocessing}
EEG data were captured using a Brain Product BrainCap MR device equipped with 64 ring-type electrodes, recording at a sampling rate of 5,000 Hz. Electrode positioning follows the standard 10-10 system, with FCz as the reference electrode and AFz as the grounding electrode. Impedance checks were conducted before each experiment to ensure that the impedance values of all electrodes remained below 5 k$\Omega$.

The continuously captured EEG data are subjected to band-pass filtering from 1 Hz to 40 Hz using an FIR filter to eliminate slow drifts and high-frequency noise. Additionally, a 50 Hz notch filter was applied to exclude line noise, followed by downsampling to 200 Hz \cite{ding2023lggnet, lee2021decoding}. Baseline correction was performed using the 1-second period before the cue. ICA decomposition was then applied to the EEG data to identify and remove components associated with eye movement (EoG) artifacts linked to the Fp1 and Fp2 electrodes. \textcolor{JamesBlue}{Additionally, muscle-related (EMG) artifacts were automatically detected and removed using MNE.} Subsequently, the continuous EEG data were segmented into epochs aligned with the markers indicating word onset. 

\section{Experiment Settings}
\label{section:Experiment_Settings}

\subsection{Training Scheme}

The training scheme for FAST and baseline models involves a subject-independent pre-training phase followed by a subject-dependent fine-tuning phase. Leave-one-subject-out (LOSO) approach is used in the pre-training phase, where the model is trained on data from all subjects except one, ensuring independence from the excluded subject. During fine-tuning, a leave-one-block-out (LOBO) \cite{zhang2021adaptive} cross-validation method is applied by dividing each subject’s data into five blocks, each containing 20 trials. In each fine-tuning iteration, four blocks serve as the training set, while the remaining block is held out for testing. This cycle is repeated until each block has been used as the test set, providing a thorough fine-tuning and evaluation across the full dataset.

Each model was trained for a fixed 200 epochs, after which training was stopped, and the accuracy of the test set was evaluated. We made sure the test dataset was only used once in the final assessment. Model performance is assessed using multi-class accuracy, F1-score, Cohen-Kappa, and AUC, present in Table \ref{tab:acc}. The equations for calculating these metrics are provided in the supplementary material.

\begin{table*}[!h]
    \centering
    \renewcommand{\arraystretch}{1.05}
    \caption{Performance comparisons in the pre-train and fine-tune phase for Covert Speech in 5 utterances of the same word}
        \textcolor{JamesBlue}{\begin{tabular}{l|l|llll}
        \hline
        \textbf{Stage} & \textbf{Model} & \textbf{Accuracy(\%)} & \textbf{F1-score} & \textbf{Cohen-Kappa} & \textbf{AUC}  \\
        \cline{1-6}
        \multirow{9}{*}{Pretrain}  
        & BIOT \cite{yang2023biot}                  & $21.8 \pm 4.6$ ***     & $0.187 \pm 0.048$ *    & $0.023 \pm 0.058$ ***  & $0.508 \pm 0.017$ ***  \\
        & EEGViT \cite{yang2023vit2eeg}             & $23.2 \pm 4.9$ **      & $0.221 \pm 0.048$      & $0.039 \pm 0.062$ **   & $0.535 \pm 0.045$ ***  \\
        & DCN \cite{schirrmeister2017deep}          & $25.0 \pm 5.7$         & $0.240 \pm 0.057$      & $0.063 \pm 0.071$      & $0.561 \pm 0.063$ ***  \\
        & EEGNet \cite{lawhern2018eegnet}           & $25.1 \pm 5.8$         & $0.238 \pm 0.060$      & $0.064 \pm 0.072$      & $0.570 \pm 0.052$ ***  \\
        & ST-Transformer \cite{song2021transformer} & $24.6 \pm 5.1$ *       & $0.228 \pm 0.060$      & $0.057 \pm 0.064$ *    & $0.557 \pm 0.048$ ***  \\
        & EEG-Conformer \cite{song2022eeg}          & $26.1 \pm 6.3$         & $0.189 \pm 0.087$      & $0.077 \pm 0.079$      & $0.608 \pm 0.083$      \\
        & EEG-Deformer \cite{ding2024eeg}           & $26.2 \pm 6.4$         & $\mathbf{0.241 \pm 0.070}$      & $0.077 \pm 0.080$      & $0.591 \pm 0.061$ *    \\
        & TSception \cite{ding2022tsception}        & $25.6 \pm 6.2$         & $0.187 \pm 0.074$ *    & $0.070 \pm 0.078$      & $0.618 \pm 0.075$      \\
        \cline{2-6}
        & FAST                                      & $\mathbf{26.9 \pm 6.9}$         & $0.221 \pm 0.091$      & $\mathbf{0.087 \pm 0.086}$      & $\mathbf{0.627 \pm 0.083}$      \\
        \hline \hline
        \multirow{9}{*}{Finetune} 
        & BIOT \cite{yang2023biot}                  & $24.0 \pm 6.5$ ***     & $0.234 \pm 0.063$ ***  & $0.050 \pm 0.081$ ***  & $0.530 \pm 0.041$ ***  \\
        & EEGViT \cite{yang2023vit2eeg}             & $24.6 \pm 5.6$ ***     & $0.244 \pm 0.056$ ***  & $0.057 \pm 0.070$ ***  & $0.558 \pm 0.059$ ***  \\
        & DCN \cite{schirrmeister2017deep}          & $26.5 \pm 6.5$ ***     & $0.264 \pm 0.066$ ***  & $0.081 \pm 0.082$ ***  & $0.577 \pm 0.072$ ***  \\
        & EEGNet \cite{lawhern2018eegnet}           & $27.2 \pm 6.5$ ***     & $0.271 \pm 0.064$ ***  & $0.089 \pm 0.081$ ***  & $0.589 \pm 0.064$ ***  \\
        & ST-Transformer \cite{song2021transformer} & $27.2 \pm 6.4$ ***     & $0.269 \pm 0.063$ ***  & $0.090 \pm 0.080$ ***  & $0.584 \pm 0.059$ ***  \\
        & EEG-Conformer \cite{song2022eeg}          & $28.6 \pm 7.0$ ***     & $0.221 \pm 0.047$ ***  & $0.107 \pm 0.087$ ***  & $0.604 \pm 0.082$ ***  \\
        & EEG-Deformer \cite{ding2024eeg}           & $30.1 \pm 7.9$ *       & $0.290 \pm 0.082$ **   & $0.126 \pm 0.099$ *    & $0.615 \pm 0.076$ **   \\
        & TSception \cite{ding2022tsception}        & $31.2 \pm 8.3$         & $0.287 \pm 0.076$ **   & $0.140 \pm 0.103$      & $0.623 \pm 0.090$ *    \\
        \cline{2-6}
        & FAST                                      & $\mathbf{34.7 \pm 10.7}$        & $\mathbf{0.340 \pm 0.108}$      & $\mathbf{0.184 \pm 0.134}$      & $\mathbf{0.662 \pm 0.097}$      \\
        \hline
        \end{tabular}}
    \label{tab:acc}
\footnotesize
\vspace{2pt}
\par Note *** denoting a $p$-value less than 0.001, ** for less than 0.01, and * for less than 0.05
\end{table*}

\begin{figure*}[!ht]
\centering
\includegraphics[width=0.9\textwidth]{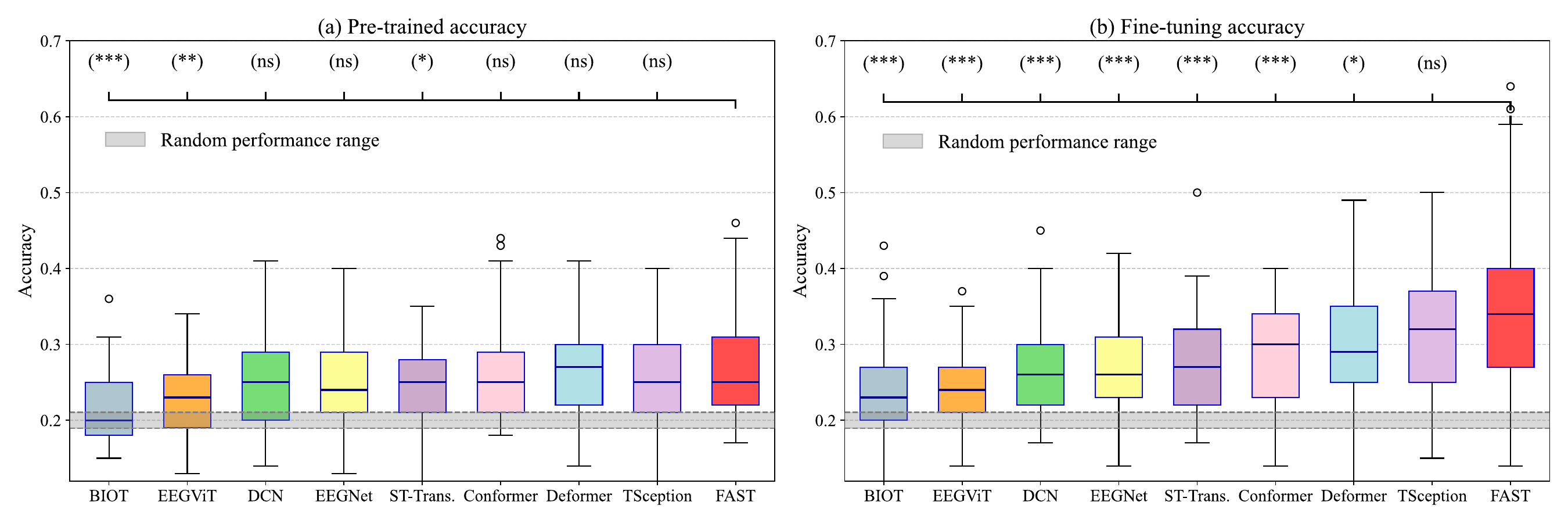}
\caption{\textcolor{JamesBlue}{Box plot illustrating the accuracy of covert speech recognition for all subjects ordered by the median accuracy: (a) Accuracy from pre-trained models; (b) Accuracy after fine-tuning. Significance levels are compared between FAST and each of the baselines, indicated with asterisks (*), where (ns) denotes p-values \textgreater 0.05. The random performance range is indicated in the gray bar.}}
\label{fig:boxplot}
\end{figure*}

\begin{figure*}[!ht]
\includegraphics[width=\textwidth]{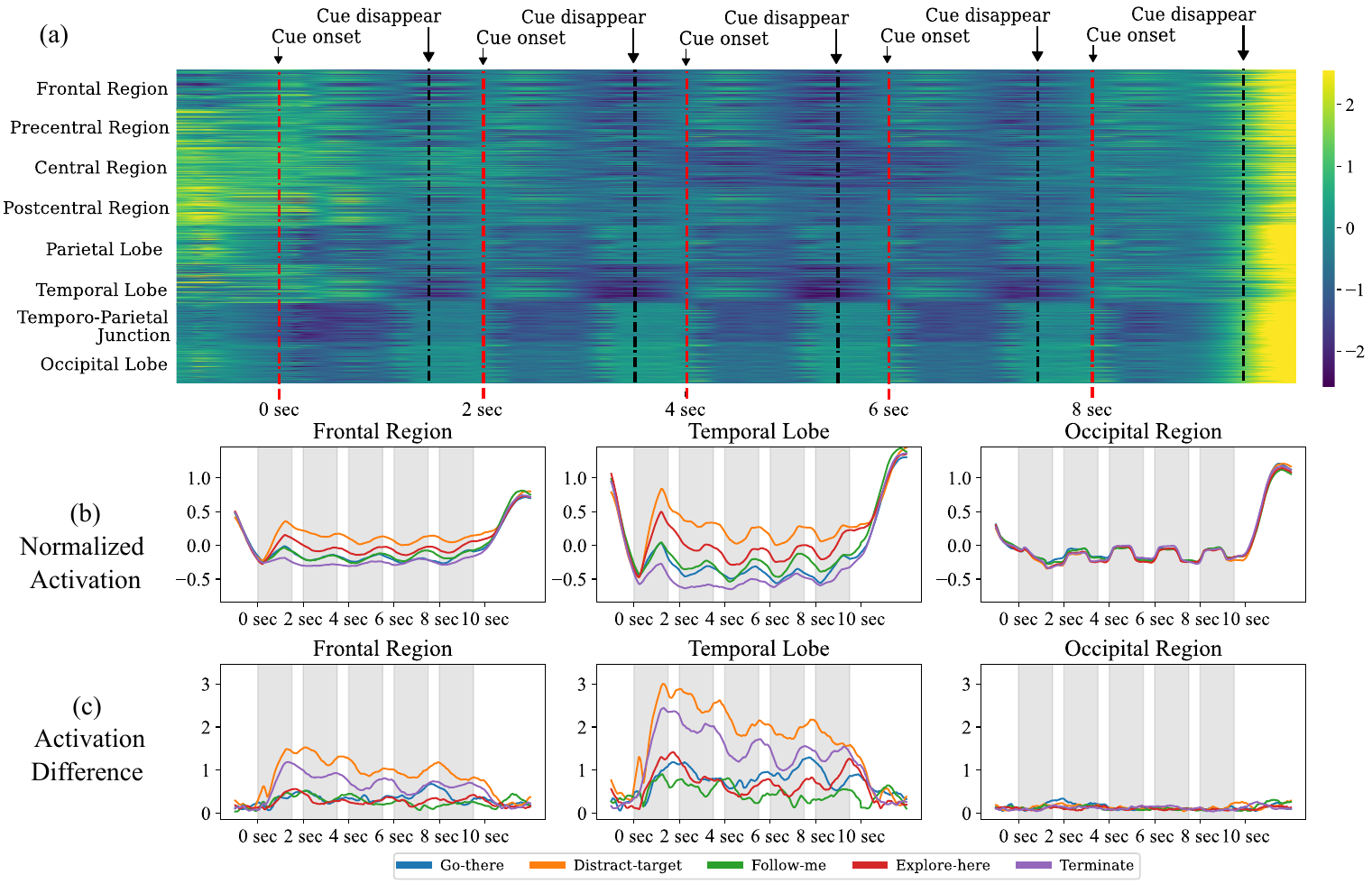}
\caption{Feature visualization of the ST on covert speech EEG data averaged across all subjects. The features were extracted from the leave-out subject after each round of leave-one-subject-out training. (a) A heatmap illustrates the features generated by the ST layers along with the corresponding time-locked features. (b) Normalized activation maps highlight the relative activation scores across the frontal, temporal, and occipital lobes. (c) Difference activation maps show the one-versus-all contrast for each region.}
\label{fig:vis}
\end{figure*}

\section{Results}
\label{section:Results_Findings}
In this section, we detail the performance of FAST for the covert speech task, We present detailed results on the 5-utterances cases in Table \ref{tab:acc}. \textcolor{JamesBlue}{We evaluate FAST against a list of commonly used models, including ST-Transformer \cite{song2021transformer}, BIOT \cite{yang2023biot}, EEGViT \cite{yang2023vit2eeg}, EEGNet \cite{lawhern2018eegnet}, DCN \cite{schirrmeister2017deep}, EEG-Conformer \cite{song2022eeg}, and TSception \cite{ding2022tsception}, as well as more recent architectures such as EEG-Deformer \cite{ding2024eeg}. All baseline models are implemented with the same pre-training and fine-tuning strategy as the proposed method.} Box plot of the accuracy visualization is shown in Figure \ref{fig:boxplot}. The chance level was determined using a binomial distribution model \cite{brown2001interval} with a random guessing probability of $p = 0.2$, and the standard deviation was calculated as $\sigma = \sqrt{p(1-p)/n}$. Using a normal approximation, the 95\% confidence interval for random accuracy was $[0.1896, 0.2104]$.

\subsection{Results and Findings}
Table \ref{tab:acc} displays the average classification accuracies, F1-score, Cohen-Kappa, AUC, and the respective standard deviations for FAST and the comparative baseline models for 5-utterances covert speech classification. Wilcoxon signed-rank test is performed between FAST and each of the baseline methods, with statistical significance levels indicated with asterisks (*), and the top-performing results highlighted in bold. The results of the test reveal that FAST significantly outperforms some baseline models during pre-training and outperforms most models except TSception during fine-tuning. 

Fine-tuning improves the performance of all models across the evaluated metrics. For instance, FAST's accuracy increases from 26.9\% in the pre-training phase to 34.7\% after fine-tuning. Similarly, baseline models such as EEG-Deformer, DCN, EEGNet, and Conformer also show notable improvements. FAST achieves the highest accuracy of 34.7\%, outperforming all baseline models. TSception, EEG-Deformer, the closest competitor, achieves 31.2\%, showing a clear advantage for FAST. FAST attains the highest F1-Score of 0.340, indicating a better balance between precision and recall than other models, including TSception (0.287). FAST records the highest Cohen-Kappa value (0.184), indicating stronger agreement compared to TSception (0.140) and EEG-Deformer (0.126). 

Figure \ref{fig:boxplot} (b) illustrates the fine-tuning accuracy distributions for each model. FAST demonstrates a higher median accuracy and reduced variability, indicating its stability and efficiency. While TSception and EEG-Deformer remain competitive. \textcolor{JamesBlue}{The reason for the lower effectiveness of baseline models is primarily due to their design limitations. These models were not originally intended to handle the complexities of speech-related data. Additionally, CNN-only models struggle with handling long input sequences, which limits their performance in our task. }

\subsection{Visualization and Analysis}
In this section, we present the visualization and analysis of the ST features extracted from covert speech EEG data. The aim is to gain a deeper understanding of the temporal and spatial activation patterns associated with covert speech processing across different brain regions.

In Fig. \ref{fig:vis}(a), visualizations of the covert speech features averaged across all subjects are shown. This averaging helps smooth out individual variability and noise, resulting in a clearer and more interpretable representation of the data. The features were derived from the left-out subject following each round of leave-one-subject-out training. The visualization is conducted by densely sliding windows across each EEG trial with a step size of 0.02 seconds. These segments are fed into pre-trained ST for feature extraction for each leave-out subject after each round of the LOSO training. Since the features output by ST have already passed through an activation layer, their values inherently represent the strength of the activation. We then averaged these features across all subjects and applied z-score normalization along the trial dimension for the plot shown in Figure~\ref{fig:vis}(a).
The normalized activation per class shown in Figure~\ref{fig:vis}(b) is computed by averaging the features on a per-class basis, with a single line representing the activation for each word and brain region. This process captures the fine-grained temporal and spatial activation during each covert speech trial. To more clearly visualize the difference in activation between words, a one-vs-all strategy is adopted, with the results illustrated in Figure~\ref{fig:vis}(c).

\textcolor{JamesBlue}{To interpret model predictions, the Integrated Gradients \cite{sundararajan2017axiomatic} method is employed, which assigns an importance score to each input feature by approximating the integral of the model's output gradients shown in Figure \ref{fig:saliency}. A detailed algorithm for generating this figure is described in the supplementary material.}

\begin{figure*}[!ht]
\centering
\includegraphics[width=1.00\linewidth]{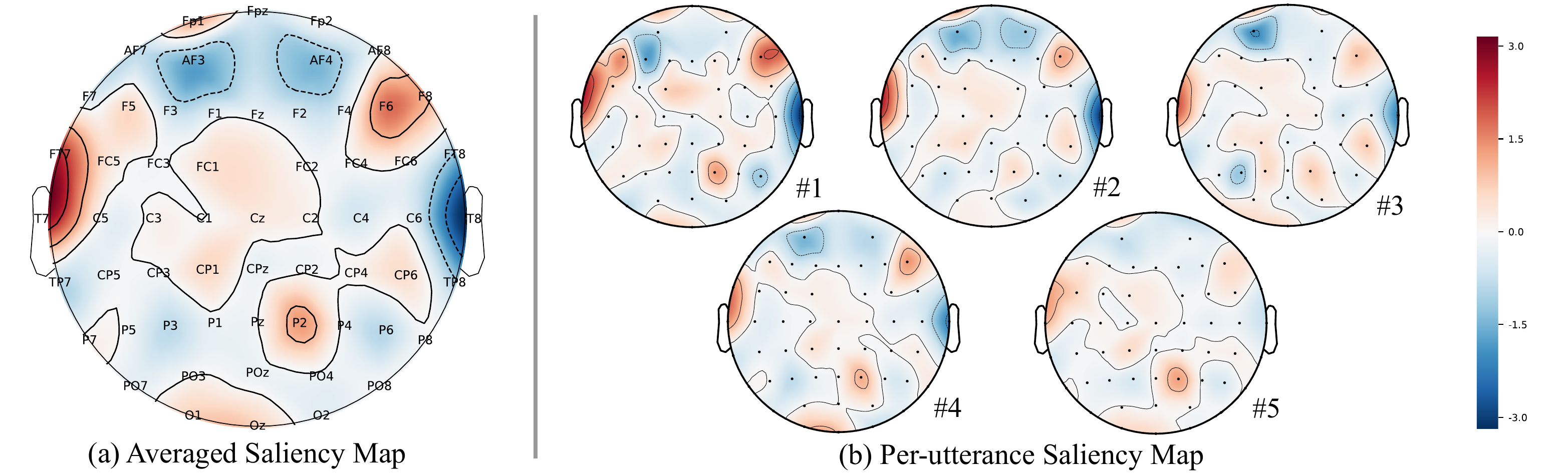}
\caption{\textcolor{JamesBlue}{EEG saliency maps during covert speech (a): Activation is prominently observed in the left hemisphere electrodes (T7, FT7, FC5, C5, F5), corresponding primarily to Broca’s area, suggesting involvement in speech motor planning and articulatory rehearsal. Additional activation is visible in electrodes F6 and F8 on the right hemisphere homologue of Broca’s area, possibly reflecting prosodic processing or cognitive control mechanisms. No activation was detected in Wernicke’s or occipital areas. (b)Activation intensity gradually decreases across successive utterances.}}
\label{fig:saliency}
\end{figure*}

\textcolor{JamesBlue}{The activation map shown in Figure~\ref{fig:vis}(a) highlights task-related responses occurring at both the onset and offset of word cues. Elevated rhythmic responses are observed in frontal and temporal regions, crucially implicated in speech processing \cite{ZHANG2024120629}, as well as in the occipital lobe, predominantly associated with visual processing \cite{vijn1992topography}. A stronger response is observed during the first cue onset. Figures~\ref{fig:vis}(b) and (c) illustrate that activation within the frontal and temporal lobes remains stable prior to \( t = 0 \) sec and becomes discriminative during the cue onset period (marked by the gray bar). Once the cue disappears at \( t = 10 \) sec, activation levels revert to baseline. Conversely, occipital lobe activation aligns with cue onset but does not show discriminative responses between the words. This finding indicates that visual information does not contribute to decoding performance. This activation pattern supports the critical role of frontal and temporal lobes \cite{blank2002speech} in covert speech processing \cite{ZHANG2024120629}, whereas occipital activation reflects visual responses to the cue rather than linguistic content.}

\textcolor{JamesBlue}{Figure~\ref{fig:saliency} illustrates the EEG saliency maps derived from the Integrated Gradients \cite{sundararajan2017axiomatic} method for the covert speech classification task. The averaged saliency map Figure~\ref{fig:saliency}(a) demonstrates pronounced neural activation predominantly localized within the left hemisphere, particularly at electrodes (T7, FT7, FC5, C5, F5), with additional notable activations at AF7 and AF3. These electrode sites correspond primarily to Broca's area and adjacent premotor regions \cite{nasios2019broca}, suggesting robust involvement in speech motor planning, articulatory rehearsal \cite{ZHANG2024120629}, and phonological processing \cite{rutten2022broca} during covert speech tasks. In addition to the left hemisphere activations, activations in the right frontal electrodes F6 and F8, which may correspond anatomically to the right hemisphere homologue of Broca’s area \cite{code1997can}. This right hemisphere activation likely indicates neural processes associated with prosodic modulation \cite{luthra2021role,buklina2018improvement,lacroix2020effects,skipper2007speech}, emotional aspects \cite{hartikainen2021emotion} of speech, or cognitive control \cite{preisig2025predictive} mechanisms during covert speech.}

\textcolor{JamesBlue}{The per-utterance saliency map depicted in Figure~\ref{fig:saliency}(b) indicates a gradual reduction in neural activation intensity across successive utterances. This diminishing pattern suggests reduced cognitive and articulatory effort with repeated covert rehearsal of speech content.}

% \textcolor{JamesBlue}{Notably, the averaged saliency map shows comparatively less activation in electrodes corresponding to Wernicke’s area, traditionally associated with language comprehension and semantic processing \cite{blank2002speech}. This absence indicates that the covert speech task mainly activates brain regions responsible for planning and rehearsing speech rather than areas involved in language comprehension \cite{binder2017current}. The per-utterance saliency map depicted in Figure~\ref{fig:saliency}(b) indicates a gradual reduction in neural activation intensity across successive utterances. This diminishing pattern suggests habituation, neural efficiency improvements, or reduced cognitive and articulatory effort with repeated covert rehearsal of speech content.}

\subsection{Ablation Study}
\begin{figure}[!ht]
\centering
\includegraphics[width=0.8\linewidth]{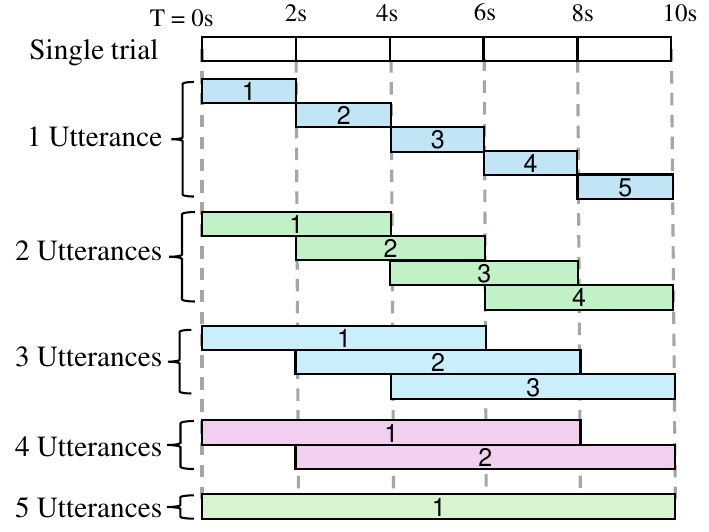}
\caption{Illustration of experiments using various utterance splits, where each 10-second trial consists of five repetitions of the same word.}
\label{fig:UtterancesSplit}
\end{figure}

To investigate whether repetitions of a word enhance decoding accuracy, the utterances split paradigm is shown in Fig \ref{fig:UtterancesSplit}. The input to the model was controlled—specifically, 2 to 10 seconds, corresponding to 1 to 5 repetitions of the word. The effect of the number of utterances of the same word on model performance is present in Table \ref{tab:acc_utterances} The results demonstrate a consistent improvement in decoding accuracy with an increasing number of utterances during both pre-training and fine-tuning stages. The improvement could be attributed to several factors. First, multiple utterance attempts reduce the noise and variability inherent in single-utterance data recording\cite{grill2006repetition,gagnepain2008spoken}, providing a more stable signal. Second, they offer richer contextual information, enabling the model to capture and learn stable cognitive patterns. Third, the aggregation of data across multiple utterances enhances the signal-to-noise ratio, leading to improved decoding accuracy.

\begin{table}[!ht]
\centering
\renewcommand{\arraystretch}{1.05}
\caption{Performance comparisons of FAST for covert speech across 1 to 5 utterances of the same word}
    \textcolor{JamesBlue}{\begin{tabular}{l|l|l}
    \hline
    N utterances & Pre-trained accuracy & Fine-tuned accuracy \\
    \hline
    1 Utterance  & $23.9\pm3.9$ & $28.4\pm6.3$ \\
    2 Utterances & $24.9\pm4.4$ & $30.5\pm7.0$ \\
    3 Utterances & $25.5\pm4.8$ & $32.6\pm7.9$ \\
    4 Utterances & $26.2\pm5.7$ & $33.3\pm8.3$ \\
    5 Utterances & $26.9\pm6.3$ & $34.7\pm10.7$ \\
    \hline
    \end{tabular}}
\label{tab:acc_utterances}
\end{table}

\begin{figure}[h]
\centering
\includegraphics[width=0.45\textwidth]{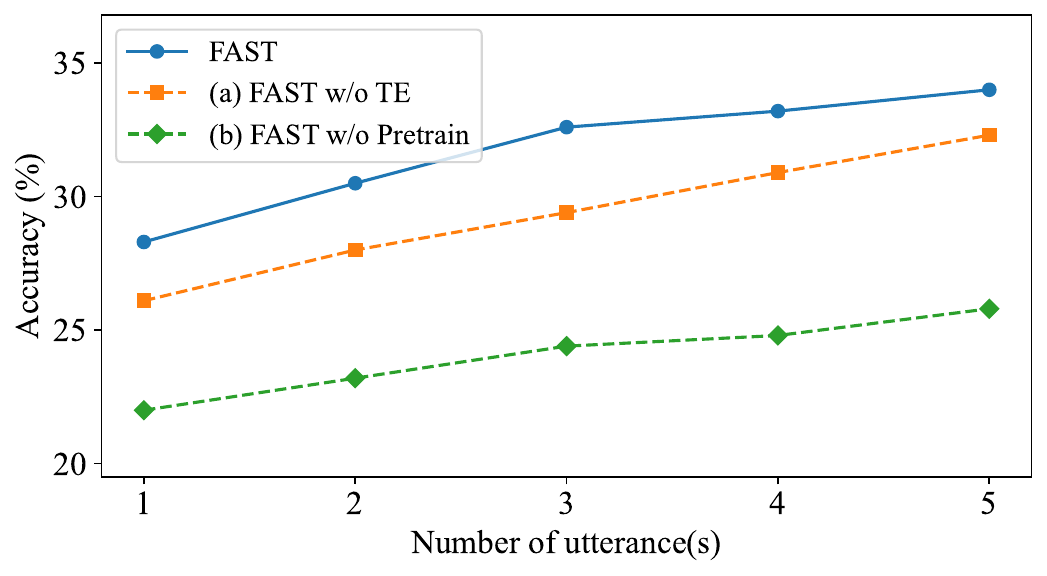}
\caption{\textcolor{JamesBlue}{The average covert speech accuracies of FAST under different ablation cases. (b) An ablation study excludes TE blocks, where the classification head is directly connected after ST. (c) Ablation study without fine-tuning, where weights are randomly initialized instead of loading from LOSO pretraining.}}
\label{fig:ablation}
\end{figure}

To investigate the individual contributions of model components and training strategies to the overall performance by selectively disabling specific elements. Two different ablation experiments are conducted. To investigate the contributions of the TE, we performed an ablation study comparing classification accuracy using only the ST and connecting the classification MLP directly after ST. The results, presented in Figure \ref{fig:ablation} (a), reveal that incorporating TE consistently improves classification accuracy across the fine-tuning phase with respect to different utterances.

To evaluate the impact of our pre-training and fine-tuning strategy, we conducted an ablation study without pre-training, where model parameters are randomly initialized rather than loaded from the pre-trained checkpoints. As shown in Figure \ref{fig:ablation} (b), the results demonstrate that the model without pre-training exhibits lower accuracy across different utterances. This suggests that pre-training plays a crucial role in enabling the model to learn generalizable representations of covert speech.

\section{Conclusion and Future Works}
\label{section:Conclusion}

In this paper, we proposed FAST an innovative covert speech architecture that incorporates ST blocks and TE blocks, demonstrating superior performance to existing baseline models. To overcome the scarcity of large-scale datasets in covert speech tasks for EEG-based BCI, we have collected a large-scale multi-utterance dataset with 57 subjects, each performing 1000 utterances per word. Analysis of extracted features from the model revealed particularly high responsiveness in the frontal temporal region of the brain, providing new insights for the understanding of neural dynamics during covert speech. We also validate the effectiveness of our algorithm on the publicly available BCI competition dataset.

\textcolor{JamesBlue}{While our study presents novel patterns related to covert speech, it has certain limitations. Notably, we included only male participants to minimize EEG variability, limiting our findings' generalizability to females. Neural responses during covert speech may differ across genders, and future research should incorporate a more diverse participant pool to enhance robustness. Additionally, expanding the dataset to different languages and demographics would further validate our approach to make a widely applicable solution for EEG-based covert speech recognition.}

\textcolor{JamesBlue}{FAST’s interpretability may offer practical benefits for other BCI applications in the future. For example, in neurofeedback, its activation patterns can help identify cognitive state-specific signatures to design personalized training protocols. In speech rehabilitation, these insights can decode neural correlates of speech, guiding the development of tailored therapeutic interventions. Overall, FAST’s detailed mapping of brain activity may provide a robust foundation for developing more targeted and effective BCI applications in the future.}

\section*{Acknowledgment}
This study is supported by the DSO National Laboratories (DSOCL21193), Singapore.

\bibliographystyle{IEEEtran}
\bibliography{Refs}

\newpage

\twocolumn
\section*{Supplementary materials}
\subsection{Comparison with Public Datasets}

\begin{table*}[h!]
\small
\centering
\caption{Comparison of datasets based on subject count, repetitions, and utterances per trial}
\begin{tabular}{l|c|c|c|c}
\hline
\textbf{Dataset} & \textbf{Classes} & \textbf{Subjects} & \textbf{Repetitions per Subject} & \textbf{Utterances per Trial} \\ \hline
BCI Competition (Imagined Speech Classification) \cite{BCICompetition2020} & 5 & 15 & 350 & 1 \\ \hline
Thinking Out Loud Dataset \cite{nieto2022thinking} & 5 & 10 & 223 (average) & 1 \\ \hline
Croetto DB \cite{coretto2017open} & 6 & 15 & 240 (average) & 1 \\ \hline
Ours & 5 & 57 & 1,000 & 5 \\ \hline
\end{tabular}
\label{tab:dataset_comparison}
\end{table*}

As shown in Table \ref{tab:dataset_comparison}, the distinctiveness of our dataset lies in the following key aspects when compared to existing freely available datasets:

\textbf{Subject Count:} Our dataset includes data from 57 subjects, which is notably larger than the referenced datasets (\cite{BCICompetition2020, nieto2022thinking, coretto2017open}). A larger subject pool improves the generalizability of the findings and allows for more robust statistical analyses, particularly in exploring inter-subject variability.

\textbf{Repetition Count Per Subject:} Each participant in our dataset completed 1,000 utterances, providing a higher volume of data per subject compared to existing datasets (\cite{BCICompetition2020, nieto2022thinking, coretto2017open}). This high repetition count enables better training and validation of machine learning models, as it provides a denser dataset for both individual-specific and generalized decoding tasks.

\textbf{Multi-utterances:} We chose to collect a multi-utterance dataset over a single-utterance dataset to ensure more stable and robust neural responses.

\subsection{Modeling Details}
In the implementation of FAST on our self-collected dataset, we used $L_t = 4$, $M=8$ and $F = 32$, for the $F_i$ feature vector, its dimension will be $8\times32=256$, and $L = 4$ for the transformer layers.
The AdamW optimizer was employed with an initial learning rate of 0.001. The learning rate schedule follows a cosine decay pattern with a warm-up phase. During the first 10 epochs, the learning rate gradually increases from 10\% to 100\%, enabling a smooth warm-up period. Following this, the learning rate decays from 100\% to 10\% over the remaining training epochs. To accelerate the pre-training phase, FP16 automatic mixed precision training was used, conserving memory and enhancing GPU throughput. For the fine-tuning phase, FP32 precision was utilized without mixed precision. We use two A100 GPUs, the experiment can be completed in approximately 5 hours.

\subsection{Metrics for Evaluation}
% \color{JamesBlue}
$$\textbf{Accuracy:} \quad \frac{TP + TN}{TP + TN + FP + FN}$$
$$\textbf{Precision (Macro):} \quad \frac{1}{C} \sum_{i=1}^{C} \frac{TP_i}{TP_i + FP_i}$$
$$\textbf{Recall (Macro):} \quad \frac{1}{C} \sum_{i=1}^{C} \frac{TP_i}{TP_i + FN_i}$$
$$\textbf{F1-score (Macro):} \quad \frac{1}{C} \sum_{i=1}^{C} \frac{2 \cdot \text{Precision}_i \cdot \text{Recall}_i}{\text{Precision}_i + \text{Recall}_i}$$
$$\textbf{Cohen's Kappa:} \quad \kappa = \frac{p_o - p_e}{1 - p_e}$$
$$p_o = \frac{1}{N} \sum_{i=1}^{N} \delta(y_i, \hat{y}_i), \quad p_e = \sum_{k=1}^{C} \left( \frac{n_k}{N} \cdot \frac{m_k}{N} \right) $$
Where \( p_o \) is the observed agreement, \( p_e \) is the expected agreement, \( n_k \) and \( m_k \) are the actual and predicted counts for class \( k \), respectively.
$$\textbf{AUC (One-vs-Rest):} \quad \text{AUC} = \frac{1}{C} \sum_{i=1}^{C} \text{AUC}_i$$
Where \( \text{AUC}_i \) is the area under the ROC curve for class \( i \), computed using one-vs-rest strategy.

\textbf{Explanation of Symbols:}
\begin{itemize}
    \item \( C \): Number of classes in the classification task.
    \item \( \delta(y_i, \hat{y}_i) \): Indicator function that is 1 if \( y_i = \hat{y}_i \) (correct prediction) and 0 otherwise.
    \item \( N \): Total number of samples in the dataset.
    \item \( n_k \): Actual number of samples belonging to class \( k \).
    \item \( m_k \): Predicted number of samples assigned to class \( k \).
    \item \( p_o \): Observed agreement - the proportion of times the predicted label matches the true label.
    \item \( p_e \): Expected agreement - the agreement expected by chance, based on class distributions.
    \item \( \text{AUC}_i \): The area under the ROC curve for class \( i \), computed using the one-vs-rest strategy.
\end{itemize}

\subsection{Visualization Details}
To interpret model predictions, the Integrated Gradients (IG) method \cite{sundararajan2017axiomatic} is employed. This method assigns an importance score to each input feature by computing the path integral of the gradients of the model's output with respect to the input, along a straight line path from a pre-defined baseline to the actual input. The IG method satisfies two key axioms: Sensitivity, which ensures that features contributing differently to the output are assigned different importance scores, and Implementation Invariance, which guarantees that two functionally equivalent models yield the same importance scores. 
The IG computation is formalized as:
\[
IG_i(x) = (x_i - x_i') \int_{\alpha=0}^1 \frac{\partial F(x' + \alpha (x - x'))}{\partial x_i} d\alpha
\]
where \(x\) is the input, \(x'\) is the baseline, \(F\) is the model's output function, and \(i\) denotes the \(i\)-th feature. This method provides an intuitive understanding of the contribution of each input feature to the model's prediction, as illustrated in Figure~\ref{fig:saliency}.

\subsection{Results on Overt Speech Decoding}
\begin{table*}[!h]
    \centering
    \renewcommand{\arraystretch}{1.05}
    \caption{Performance comparisons Covert/Overt speech decoding for 5 utterances}
        \textcolor{JamesBlue}{\begin{tabular}{l|l|llll}
        \hline
        \textbf{Stage} & \textbf{Condition} & \textbf{Accuracy} & \textbf{F1-score} & \textbf{Cohen-Kappa} & \textbf{AUC}  \\
        \cline{1-6}
        \multirow{2}{*}{Pretrain}
        & Covert  & $26.9\pm6.3$ & $0.221\pm0.091$ & $0.087\pm0.086$ & $0.627\pm0.083$ \\
        & Overt   & $30.1\pm8.8$ & $0.234\pm0.113$ & $0.126\pm0.111$ & $0.681\pm0.096$ \\
        \hline \hline
        \multirow{2}{*}{Finetune} 
        & Covert  & $34.7\pm10.7$ & $0.340\pm0.108$ & $0.184\pm0.134$ & $0.662\pm0.097$ \\
        & Overt   & $51.3\pm12.9$ & $0.523\pm0.104$ & $0.496\pm0.089$ & $0.721\pm0.058$ \\
        \hline
        \end{tabular}}
    \label{tab:diff_covert_overt}
\footnotesize
\end{table*}

the decoding accuracy for overt speech is higher compared to covert speech, as seen in Table \ref{tab:diff_covert_overt}. This difference can be attributed to two key factors. First, overt speech likely produces stronger and more consistent neural activity due to the physical act of speaking, which enhances the reliability of the decoding process. Second, the muscle movements involved in overt speech may generate electromyographic signals, providing additional information that aids in classification.

\subsection{Ablation on Brain Areas Partitioning Configurations}
\begin{figure*}[t]
\centering
\includegraphics[width=0.9\linewidth]{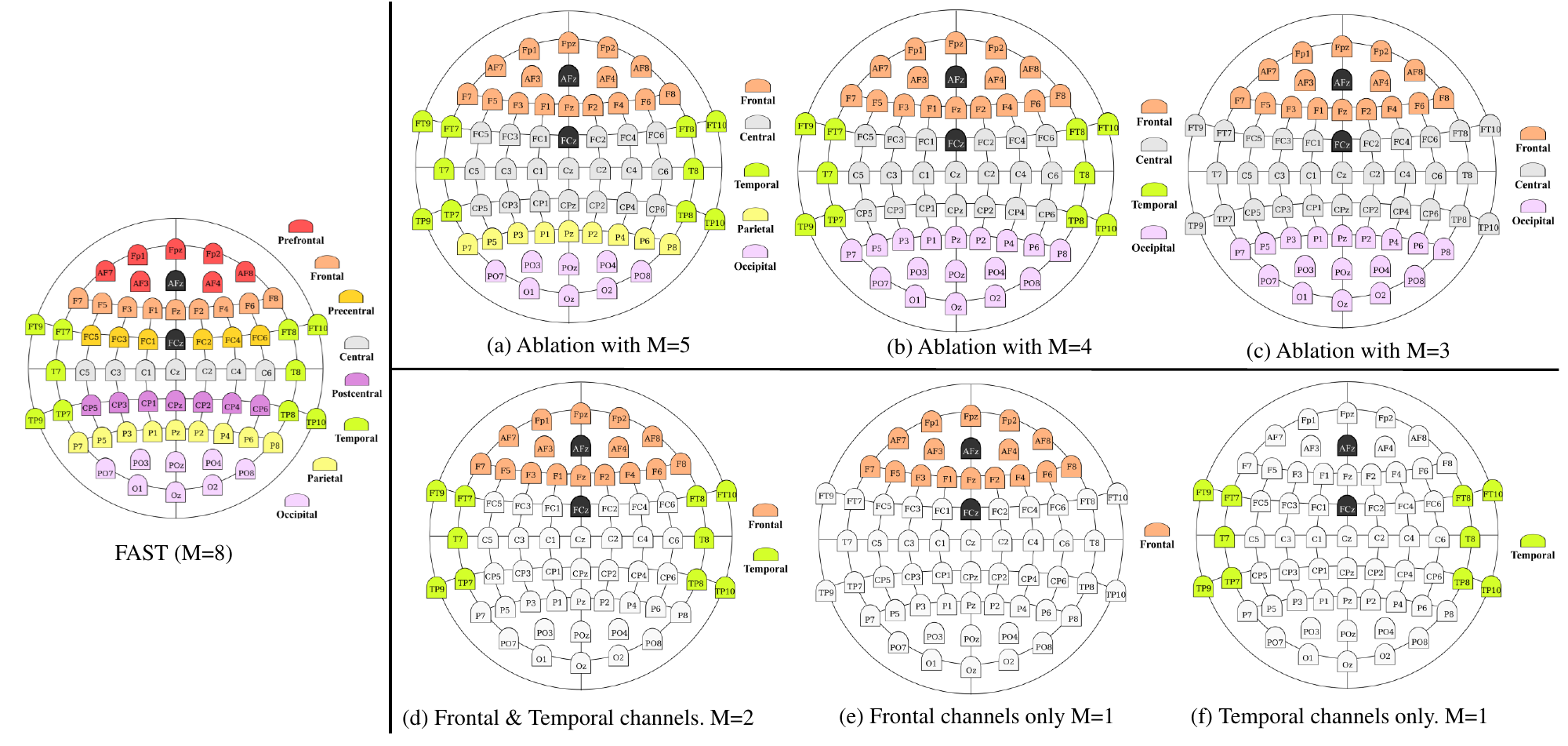}
\caption{illustration of different M and configurations. (a) M=5 by merging the electrodes from the prefrontal into frontal, precentral, and postcentral central; (b) M=4 by additionally merging the parietal into occipital; (c) M=3 by only keeping frontal, central, and occipital. (d) using both Frontal and Temporal electrodes, (e) using Frontal electrodes only, and (f) using Temporal electrodes only.}
\label{supp:Zone_Ablation}
\end{figure*}

In addition to the M=8 in the manuscript, here we add the following configurations: (a) M=5 by merging the electrodes from prefrontal into frontal, precentral, and postcentral into central; (b) M=4 by additionally merging parietal into occipital; (c) M=3 by only keeping the frontal, central and the occipital. The performance is shown in the Table \ref{tab:diff_M} (a)(b)(c). The performance metrics generally decrease as the number of regions M decreases from FAST to conditions (a)(b) and (c). 

To examine the impact of using only speech-related electrodes, three additional ablation studies were conducted: (d) using both Frontal and Temporal electrodes, (e) using Frontal electrodes only, and (f) using Temporal electrodes only. As shown in Table \ref{tab:diff_M} (d)(e)(f), the baseline FAST model (M=8) outperforms all ablated configurations in both pre-train and finetune stages. Among the ablations, temporal electrodes were shown more impactful during finetuning as shown in (c). 

\begin{table*}[!h]
    \centering
    \renewcommand{\arraystretch}{1.05}
    \caption{Performance comparisons with different M and configurations}
        \textcolor{JamesBlue}{\begin{tabular}{l|l|llll}
        \hline
        \textbf{Stage} & \textbf{Model} & \textbf{Accuracy} & \textbf{F1-score} & \textbf{Cohen-Kappa} & \textbf{AUC}  \\
        \cline{1-6}
        \multirow{7}{*}{Pretrain}
        & FAST (M=8)                   & $26.9\pm6.9$ & $0.221\pm0.091$ & $0.087\pm0.086$ & $0.627\pm0.083$ \\
        \cline{2-6}
        & (a) M=5                    & $25.5\pm5.2$ & $0.163\pm0.067$ & $0.069\pm0.065$ & $0.633\pm0.073$ \\
        & (b) M=4                    & $25.6\pm5.5$ & $0.168\pm0.065$ & $0.071\pm0.068$ & $0.634\pm0.076$ \\
        & (c) M=3                    & $25.9\pm5.6$ & $0.178\pm0.074$ & $0.074\pm0.070$ & $0.637\pm0.074$ \\
        & (a) M=2 Frontal + Temporal & $26.9\pm5.5$ & $0.195\pm0.072$ & $0.087\pm0.069$ & $0.638\pm0.072$ \\
        & (b) M=1 Frontal Only       & $26.4\pm5.1$ & $0.180\pm0.068$ & $0.080\pm0.064$ & $0.637\pm0.071$ \\
        & (c) M=1 Temporal Only      & $27.4\pm6.4$ & $0.215\pm0.079$ & $0.092\pm0.080$ & $0.642\pm0.072$ \\
        \hline \hline
        \multirow{7}{*}{Finetune} 
        & FAST (M=8)                   & $34.7\pm10.7$ & $0.340\pm0.108$ & $0.184\pm0.134$ & $0.662\pm0.097$ \\
        \cline{2-6}
        & (a) M=5                    & $33.1\pm7.0$ & $0.291\pm0.073$ & $0.164\pm0.087$ & $0.645\pm0.080$ \\
        & (b) M=4                    & $32.7\pm7.8$ & $0.288\pm0.082$ & $0.159\pm0.097$ & $0.643\pm0.081$ \\
        & (c) M=3                    & $33.4\pm7.2$ & $0.293\pm0.072$ & $0.167\pm0.090$ & $0.652\pm0.077$ \\
        & (d) M=2 Frontal + Temporal & $33.7\pm8.1$ & $0.299\pm0.083$ & $0.172\pm0.102$ & $0.652\pm0.083$ \\
        & (e) M=1 Frontal Only       & $33.8\pm8.4$ & $0.289\pm0.087$ & $0.172\pm0.105$ & $0.649\pm0.078$ \\
        & (f) M=1 Temporal Only      & $34.2\pm8.7$ & $0.316\pm0.087$ & $0.178\pm0.109$ & $0.651\pm0.078$ \\
        \hline
        \end{tabular}}
    \label{tab:diff_M}
\footnotesize
\end{table*}

\subsection{Additional Public Dataset}

In addition to the self-collected dataset, to validate the effectiveness of FAST, we included the publicly available multi-class imagined speech classification dataset from the 2020 International BCI Competition \cite{BCICompetition2020}, this dataset consists of EEG recordings from 15 subjects, aged 20-30 years old. The subjects were instructed to imagine the silent pronunciation of five conversational words/phrases: 'hello,' 'help me,' 'stop,' 'thank you,' and 'yes.' given the word as if they were performing real speech, without moving any articulators or making the sound. Each subject participated in 70 trials per class, yielding a total of 350 trials (60 trials per class for training and 10 trials per class for validation). EEG signals were recorded from 64 electrodes following the 10-20 international system. Ground and reference electrodes were placed at Fpz and FCz.

For this dataset, the limited subject count (15 subjects) led to subject-independent pre-training being ineffective for both FAST and baseline models. Therefore, we proceeded directly with subject-dependent training, performing a 5-fold cross-validation on each subject's 350 trials.

\subsection{Results on public dataset}

\begin{table}[!h]
\centering
\caption{Performance comparisons on 2020 International BCI Competition Dataset, Track 3: Imagined Speech Classification}
\label{tab:T3_acc}
\begin{tabular}{l|l}
\hline
\textbf{Method} & \textbf{Accuracy (1 utterance) } \\ \hline
LSTM                                      & 36.5 ± 4.8 ***\\ 
GRU                                       & 38.3 ± 4.7 ***\\ 
DeepConvNet \cite{schirrmeister2017deep}  & 33.2 ± 5.8 ***\\ 
EEGNet \cite{lawhern2018eegnet}           & 40.6 ± 6.2 ***\\ 
EEG-Conformer \cite{song2022eeg}          & 43.3 ± 5.1 ***\\
Dong-Yeon et al.\cite{lee2021decoding}    & 48.1 ± 3.6 (NA) \\ 
TSception \cite{ding2022tsception}        & 52.2 ± 8.8 *** \\
\hline
FAST                                      & \textbf{54.8 ± 9.1} \\ \hline
\end{tabular}
\vspace{2pt}
\par Note *** denoting a $p$-value less than 0.001, ** for less than 0.01, and * for less than 0.05
\end{table}

Table \ref{tab:T3_acc} provides the classification accuracies and standard deviations across various models on the BCI competition dataset, the dataset only contains a single utterance per trial. The table also presents the results of the Wilcoxon signed-rank test, with significant p-values marked by asterisks. The individualized accuracy for \cite{lee2021decoding} is unavailable and is indicated as (NA). FAST achieved the highest average accuracy at 54.8\% with a standard deviation of 9.1\%. Among the models evaluated, EEG-Conformer, \cite{lee2021decoding}, and TSception also demonstrated competitive performance, achieving average accuracies of 43.38\%, 48.10\%, and 52.26\% respectively. The FAST model outperformed all other approaches.

\end{document}